\documentclass[11pt]{article}
\usepackage{amsmath,amssymb,times,graphicx,mathdots}
\begin{document}
\pagestyle{plain}
\hsize = 6. in 				
\vsize = 8.5 in		
\hoffset = -0.6 in
\voffset = -0.5 in

\title{The Dynamics of Zeroth-Order Ultrasensitivity:
A Critical Phenomenon in Cell Biology}

\author{Qingdao Huang\footnote{huangqd@jlu.edu.cn}\\[5pt]
College of Mathematics, Jilin University\\
Changchun 130012, PRC
\\[8pt]
and\\[8pt]
Hong Qian\footnote{qian@amath.washington.edu}\\[5pt]
Department of Applied Mathematics,
University of Washington\\
Seattle, WA 98195, USA
}

\maketitle

\begin{abstract}

    It is well known since the pioneering work of
Goldbeter and Koshland [Proc. Natl. Acad. Sci. USA, vol. 78, pp.
6840-6844 (1981)] that cellular phosphorylation- dephosphorylation
cycle (PdPC), catalyzed by kinase and phosphatase under saturated
condition with zeroth order enzyme kinetics, exhibits
ultrasensitivity, sharp transition.  We analyse the dynamics aspects
of the zeroth order PdPC kinetics and show a critical slowdown akin
to the phase transition in condensed matter physics. We demonstrate
that an extremely simple, though somewhat mathematically
``singular'' model is a faithful representation of the
ultrasentivity phenomenon.  The simplified mathematical model will
be valuable, as a component, in developing complex cellular
signaling netowrk theory as well as having a pedagogic value.


\end{abstract}

\section{Introduction}

    Cellular informations are stored in the genome; cellular
functions, on the other hand, are carried out by biochemistry.  The
communications within and between cells are based on biochemical
reactions involves enzyme molecules \cite{rob_book}.  In particular,
reversible chemical modification of enzymes, via phosphorylation-
dephosphorylation cycle (PdPC), is one of the most widely used
mechanisms in cellular signaling and regulation
\cite{fischer_kreb_55}.  In its simplest, canonical form, an enzyme
$E$ can be phosphorylated to become $E^*$, and then dephosphorylated
to be back to $E$.  These two reactions are catalyzed by their
respective enzymes called kinase and phosphatase.  Biologically, the
$E$ form of the enzyme is inactive while the $E^*$ form is active.
Hence, a change in the concentation of the kinase (or of the
phosphatase) leads to a change in the activity of the enzyme
\cite{qian_arpc_07,qian_book}.

    Mathematical modeling of the PdPC kinetics began
in the late 1970s \cite{stadt_chock_77}.  In 1981,
Goldbeter and Koshland discovered a surprising result:
If both kinase and phosphatase, acting as enzymes for
their respective reactions, are operating under a saturated
condition, then the change of $E\rightarrow E^*$ can be
extreme sharp in response to a change in the kinase
concentration.  They called this phenomenon zeroth-order
ultrasensitivity \cite{gk_pnas_81}.  Before that work,
biochemists had believed a sharp, sigmoidal transition is
the signature of only enzyme systems with multiple
subunits or monomer with  multiple binding sites;
sigmoidal shaped transition is associated with equilibrium
cooperativity.  What had not been fully appreciated until recent
years is that the PdPC is not a closed equilibrium system, but
in fact an open-chemical system \cite{qian_arpc_07}.

    There is now a sizable literature on
ultrasensitive PdPC and related biochemical reactions.
See \cite{gunawardena09,huang09} for two very recent reports.
But all the studies, as far as we know, have been
limited to the steady state analysis, with the exception
of \cite{wang_02}.   The present
paper reports a study of the time-dependent behavior
of ultrasensitive PdPC.  We discover that in addition
to the sharp transition, the kinetic of ultrasensitive PdPC
exhibit a critical slowdown, i.e., the time to reach
steady state tends to infinit when the system
approaches the the critical transition point.
In a separated study, we have also discovered that
at the critical point of transition, the ultrasensitive
PdPC exhibits large fluctuations \cite{qian_cooper_08,qian_bpc_03}.

    Putting all these together, this is a reminiscent
of the critical phenonemon
in physics \cite{domb_book,stanley_rmp_99}.  The classical
phase transition problems in condensed matter physics,
however, are equilibrium phenomena.  The ultrasensitivity
discussed in the present work is a nonequilibrium
phenomenon.  PdPC is a driven chemical reaction system
\cite{qian_arpc_07}.

\section{Phosphorylation-Dephosphorylation Cycle Kinetics}

The canonical kinetic system for the PdPC is
\cite{gk_pnas_81,bnk_04,gunawardena}
\begin{equation}
\label{the_rxn}
    E + K \overset{k_{11}}{\underset{k_{-11}}
        {\rightleftharpoons}} EK
    \overset{k_{12}}{\longrightarrow} E^* + K, \ \ \
    E^* + P \overset{k_{21}}{\underset{k_{-21}}
        {\rightleftharpoons}} E^*P
    \overset{k_{22}}{\longrightarrow} E + P,
\end{equation}
in which $K$ and $P$ represent kinase and phosphatase.
We follow the Michaelis-Menten approach to model
the enzyme reactions \cite{murray_book,qian_book}.
If the amount of kinase is significantlly smaller
than that of $E$, and the amount of phosphatase is
significantly smaller than that of $E^*$, then according
to the Michaelis-Menten theory of enzyme kinetics, we
have \cite{murray_book,qian_book}
\begin{equation}
\label{the_eq}
    \frac{dc_{E^*}}{d\tau} = \frac{V_1 c_E }{K_{M1}+c_E}
    - \frac{V_2 c_{E^*}}{K_{M2}+c_{E^*}},
\end{equation}
in which the total concentration $c_E+c_{E^*}=c_T$ is
a constant.  Here $c_E$ and $c_{E^*}$ are the
concentrations of $E$ and $E^*$ respectively.
The parameters in Eq. (\ref{the_eq}) are related to
the rate constants in (\ref{the_rxn}):
\begin{equation}
    K_{Mi} = \frac{k_{-i1}+k_{i2}}{k_{i1}}, \ \ (i=1,2); \ \
    V_1 = k_{12}[K]_{tot}, \ \ V_2 = k_{22}[P]_{tot},
\end{equation}
in which $[K]_{tot}$ and $[P]_{tot}$ are the total concentrations
of kinase and phosphatase, respectively.

    Equation (\ref{the_eq}) can be rigorously
justified via singular perturbation theory \cite{murray_book,qian_book}.
It is valid for all times, as long as one can neglects the extremely
fast kinetics.  Eq. (\ref{the_eq}) in fact is exact for finding the
steady state of the reaction system in (\ref{the_rxn}). The
approximation comes from $c_E+c_{E^*}=c_T$ since it neglects the
concentrations of $EK$ and $E^*P$.

     We introduce nondimensionalized variables
\begin{equation}
    x = \frac{c_{E^*}}{c_T}, \ \ t = \frac{V_2}{c_T}\tau,
\end{equation}
then we have
\begin{equation}
\label{the_eq1}
    \frac{dx}{dt} = \frac{\theta (1-x)}
            {K_1+1-x}
        - \frac{x}{K_2+x},
\end{equation}
where
\begin{equation}
   \theta = \frac{V_1}{V_2}=\frac{k_{12}[K]_{tot}}{k_{22}[P]_{tot}}, \ \
    K_1 = \frac{K_{M1}}{c_T}, \ \
    K_2 = \frac{K_{M2}}{c_T}.
\end{equation}
Eq. (\ref{the_eq1}) is the starting point of our
current study.   The steady state $x^*$ as a function of
$\theta$ and $K$s is well understood \cite{gk_pnas_81,qian_book}.
We investigate its time-dependent behavior.

\section{A Simple Mathematical Model for Ultrasensitivity}

    The two terms on the right-hand-side of Eq. (\ref{the_eq1})
represent the {\em Michaelis-Menten rate law} of enzyme
catalyzed reactions, for the kinase and the phosphatase,
respectively.  Taking the term $r=x/(K_2+x)$ as example, it
has the celebrated ``hyperbolic'' saturation form, also known
as double reciprocal since $1/r$ is a linear function of $1/x$
\cite{qian_book}.  Therefore, the rate $r$ as a function of
the substrate concentration $x$ is linear, i.e., first-order,
when $K_2\gg x$, but zeroth order when $K_2\ll x$.
When an enzyme is operating under zeroth order condition, the
catalyzed reaction rate is independent of the substrate
concentration.

    The important discovery of Goldbeter and Koshland
\cite{gk_pnas_81} is that when both kinase and phosphatase
in (\ref{the_rxn}) are operating under zeroth order condition,
the steady state concentration of phosphorylated enzyme, i.e.,
$x$ has a very sharp response to the $\theta$, the activation
parameter.

\begin{figure}[t]
\[
\includegraphics[height=5in,angle=-90]{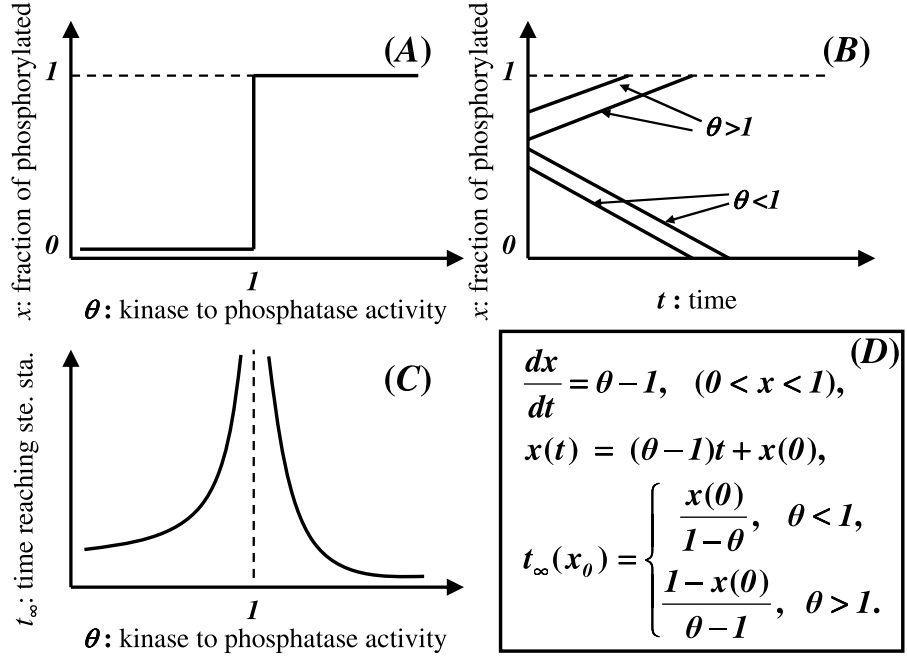}
\]
\caption{The simple zeroth-order, ultrasensitivity transition in a
nutshell: $x$ represents in the fraction of enzyme being
phosphorylated; and $\theta$ is the ratio of the activities of the
kinase and the phosphatase. (A) The activation curve, steady state
$x$ as a function of $\theta$, shows a sharp transition. (B) The
time-dependent $x(t)$ is a linear function of time $t$, not
exponential as usually expected from first-order chemical kinetics.
It reaches steady state in finite time $t_{\infty}$. (C) The time to
reach the steady state, $t_{\infty}$, exhibit a critical slowdown
near the midpoint of the transition, when $\theta=1$. (D) summarizes
the key mathematical results associated with Eq. \ref{unper_eq}. }
\label{fig1}
\end{figure}

    One can solve the steady state of Eq. (\ref{the_eq1}),
as a root of a quadratic equation, then let $K_1$ and $K_2\rightarrow
0$.  This is the standard method of attack.  However, if one lets
$K_1=K_2=0$, one immediately obtains
\begin{equation}
\label{singular_eq}
    \frac{\theta (1-x)}{1-x} = \frac{x}{x}.
\end{equation}
Being mathematically careful with Eq. (\ref{singular_eq}),
one has
\begin{equation}
    x=0; \ \textrm{ or } \
    x=1; \  \textrm{ or} \ \
    \theta = 1, \  \textrm{ if } x\neq 0, x\neq 1.
\end{equation}
Noting that $x$ is a monotonic increasing function of $\theta$ in
Eq. (\ref{the_eq1}), we have steady state $x$ as a function
of $\theta$ given in Fig. \ref{fig1}A.

    This way of setting $K_1=K_2=0$ is in fact the
spirit of {\em singular perturbation}
method \cite{murray_book}.  Following the same
approach and treating $K_1=K_2=\epsilon$ as a
small parameter, we have the unperturbed
equation
\begin{equation}
\label{unper_eq}
    \frac{dx}{dt} = \theta - 1,  \ \ \
    (0< x< 1)
\end{equation}
The time-dependent solution to the equation is
\begin{equation}
\label{the_td_sol}
    x(t) = (\theta-1)t + x(0), \ \
    \textrm{ for } \ \ 0\le x\le 1.
\end{equation}
This is shown in Fig. \ref{fig1}B.
The most interesting insight from this simple
solution is that the time reaching the steady state,
$t_{\infty}$, is finite when $\theta\neq 1$.
Furthermore, the time
\begin{equation}
\label{the_time}
    t_{\infty} = \left\{\begin{array}{lc}
        \frac{x(0)}{1-\theta} & \theta < 1,
        \\[7pt]
        \frac{1-x(0)}{\theta-1}  & \theta > 1.
\end{array}\right.
\end{equation}
This is shown in Fig. \ref{fig1}C.

    The $\theta=1$ is known as the mid-point of
transition from unphosphorylated $E$ to the phosphorylated $E^*$.
It is the ``critical point'' of the transition.  A
shape transition like in Fig. \ref{fig1}A is usually identified as
cooperative phase transition in condensed matter physics.   The
result on the time in Eq. (\ref{the_time}) and Fig. \ref{fig1}C
further collaborate this analogue: One of the characteristics of
phase transition is the ``critical slowdown'' in its dynamics
\cite{stanley_rmp_99}.   In the theory of phase transition, critical
slowdown and shape transition are also intimately related to large
fluctuations \cite{stanley_rmp_99}.  This last aspect of PdPC has
been discussed in \cite{qian_bpc_03,qian_cooper_08}.

    The easiness and transparency of Eq. (\ref{unper_eq})
makes it very attractive as a simple model for ultrasensitivity. The
remaining of the paper is to provide a more rigorous justification
of the results obtained from Eq. (\ref{unper_eq}), summarized in
Fig. \ref{fig1}D, in terms of a perturbation expansion approach.
More specifically, we shall try to provide perturbation corrections
to the results in Figs. \ref{fig1}B and \ref{fig1}C.

\section{Exact Solution to PdPC Kinetics Eq. (\ref{the_eq1})}

    To justify the above very simple model for ultrasensitivity,
one can in fact solve exactly the nonlinear differential equation
in (\ref{the_eq1}).  For simplicity, we shall assume $K_1=K_2=K$.
The Eq. (\ref{the_eq1}) can be re-written as
\begin{equation}
\label{the_eq2}
    \frac{dx}{dt} = \frac{(1-\theta)x^2+(\theta-\theta K
        -K -1)x+\theta K}{(K+1-x)(K+x)},
\end{equation}
the numerator of whose right-hand-side has two roots
$x_1$ and $x_2$:
\begin{equation}
    x_{1,2} = \frac{-\theta+\theta K+K+1\pm\sqrt{
    (1+2K)(\theta-1)^2+K^2(\theta+1)^2}}{2(1-\theta)}.
\end{equation}
One of the $x_1$ and $x_2$ is $\in [0,1]$, and it is the
steady state of the Eq. (\ref{the_eq1}).  Let us denote it
by $x^*$,  then the other one is $x' = \theta K/[x^*(1-\theta)]$.
If $0\leq\theta\leq 1$, $x_2\in [0,1]$, and it is the steady state of
the Eq. (\ref{the_eq1}). We also see that $K\rightarrow
0,x_2\rightarrow 0$. If $\theta\geq 1$, $x_1\in [0,1]$, and we have
$K\rightarrow 0,x_1\rightarrow 1$.

    Applying separation of variables and the method of
partial fraction to Eq. (\ref{the_eq2}), we have the solution
to the Eq. (\ref{the_eq1})
\begin{equation}
    t =\frac{1}{1-\theta}
    \left\{ x(0)-x(t) + \frac{K}{1-\theta}\ln\left[
        \left(\frac{x(t)-x^*}{x(0)-x^*}\right)^A
        \left(\frac{x(t)-x'}{x(0)-x'}\right)^B \right]
        \right\}.
\label{the_sol}
\end{equation}
in which,
\begin{eqnarray*}
&&  A = \frac{1+K-\theta K - (1+\theta)x^*}{x^*-x'},
\\[7pt]
&&  B = -\frac{1+K-\theta K- (1+\theta)x'}{x^*-x'}.
\end{eqnarray*}

    From Eq. (\ref{the_sol}), it is immediately clear that
if $K=0$, then $x = (\theta-1)t + x(0)$.  This is the result in
Eq. (\ref{the_td_sol}).  Furthermore, we can compute the ``time to
steady state''.   If $K\rightarrow 0$, $t\rightarrow (x_0-x(t))/(1-\theta)$.
Hence for $x(t_{\infty})=0$, $t_{\infty}=\frac{x(0)}{1-\theta}$,
and for $x(t_{\infty})=1$, $t_{\infty}=\frac{x(0)-1}{1-\theta}$.
These are the results in Eq. (\ref{the_time}).

    For $K\neq 0$, mathematically exponential decay takes
infinite time to actually reach the steady state.  To give a
measure of the time, however, we consider the time from
$x(0)=x^*+0.01$ to $x(t_{\infty})=x^*+0.001$, for $0\leq \theta< 1$.
Similarly, for $\theta > 1$, we consider the time from $x(0)=x^*-0.01$
to $x(t_{\infty})=x^*-0.001$.

\begin{figure}[t]
\[
\includegraphics[height=4in,width=3.25in,angle=-90]{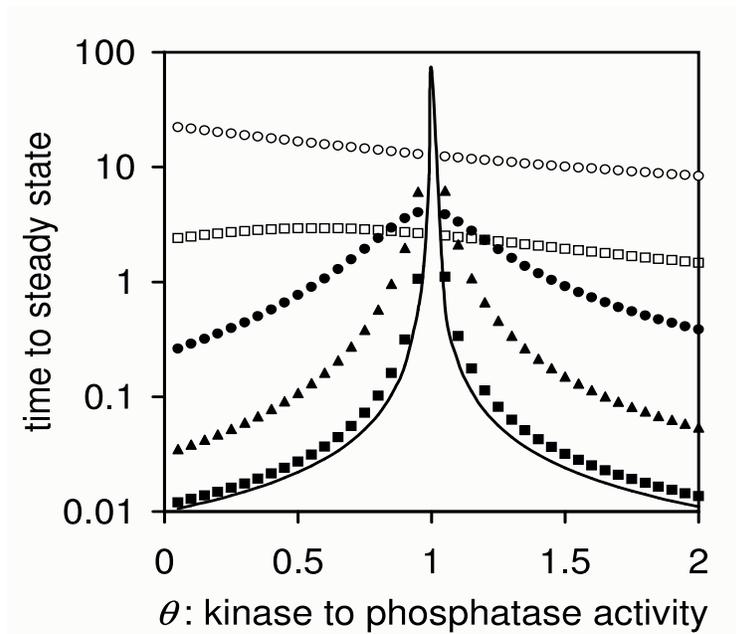}
\]
\caption{Time to reach steady state ($x^*)$, defined as from initial
value $x(0)=x^*\pm 0.01$ to $x(t_{\infty})=x^*\pm 0.001$. $\pm$
corresponds to $\theta<1$ and $>1$, respectively. The values for $K$
are 10 (open circles), 1 (open squares), 0.1 (filled circles), 0.01
(filled triangle), and 0.001 (filled square).  The solid line is
from Eq. (\ref{the_time}).} \label{fig2}
\end{figure}

    Fig. \ref{fig2} shows the $t_{\infty}$ as a function
of $\theta$, with various $K$ values.   It is shown that with
sufficiently small $K$, the result in Eq. (\ref{the_time}) is indeed
valid; There is a critical slowdown at the $\theta=1$ for zeroth
order ultrasensitivity with small $K$.

\begin{figure}[t]
\[
\includegraphics[height=3.5in,angle=-90]{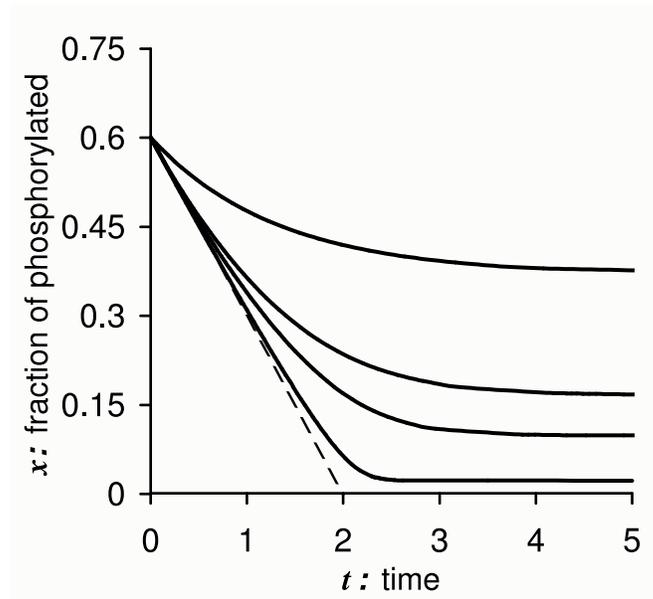}
\]
\caption{Time dependence of PdPC kinetics.  $x$ represents
the fraction of substrate enzyme being phosphorylated ($c_{E^*}/c_T)$.
All curves have initial value $x(0)=0.6$ and $\theta=0.7$.
Different curves correspond to
different $K$ values: solid lines from top to bottom
correspond to $K = 1$, 0.1, 0.05, and 0.01, respectively.
The dashed line is from Eq. (\ref{the_td_sol}). With
diminishing $K$, the curves asymptotically approaches to
the dashed line. }
\label{fig3}
\end{figure}

    We can also consider the time-dependent solution given
in Eq. (\ref{the_sol}), $x(t)$.  Fig. \ref{fig1}B
shows simple linear functions of time, each of which reaches steady
state in finite time.   To better understand this observation,
Fig. \ref{fig3} shows that the $x(t)$
asymptotically approaches to Eq. (\ref{the_td_sol})
with diminishing $K$.

\section{Discussion}

	In developing a mathematical model for a scientific
problem, one should always follow Occam's Razor which states, in
Albert Einstein's words ``as simple as possible, but not simpler.''
The present paper contains two results:  First, it proposes
an extremely simple equation for the zeroth-order PdPC, and provides a
mathematical proof that the simple model captures all the essential
features of the well-accepted, but more complex model.
Second, we analyzed the time-dependent behavior of zeroth-order
ultrasensitivity and demonstrated a critical slowdown at the
transition point; Thus it is a critical phenomenon.

    Critical phenomena are the emergent properties of
cooperative processes \cite{domb_book,hct,stanley_rmp_99}.
Calling the zeroth-order
ultrasensitivity a critical phenomenon, thus, begs an answer to
the question ``what is the cooperativity in ultrasensitivity''?
Answer to this deeper question is outside the scope of the
present paper; we refer the readers to two recent papers
on the so-called {\em temporal cooperativity}
\cite{qian_cooper_08,gehao_08}. As a matter of fact, a
nonequilibrium critical phenomenon in a driven system
emerging from a cooperativity in {\em time} seems to be a
novel concept offered by the present mathematical model of the
biological system.

\section{Acknowledgement}

	We thank Professors Chao Tang, Zhi-Xin Wang
and Jianhua Xing for interesting discussions.
The first author is a receipant of a scholarship from the
China Scholarship Council.

\end{document}